\documentclass[final,5p,twocolumn]{autart}
\usepackage{graphicx}
\usepackage{mathtools}
\usepackage{txfonts}

\usepackage{enumitem}
\setitemize{nolistsep}
\setenumerate{nolistsep}
\usepackage{todonotes}

\newtheorem{conv}[thm]{Convention}

\newcommand{\R}{\mathbb{R}}
\newcommand{\N}{\mathbb{N}}
\renewcommand{\P}{\mathcal{P}}

\newcommand{\abs}[1]{\left\lvert{#1}\right\rvert}
\newcommand{\norm}[1]{\left\lVert{#1}\right\rVert}
\newcommand{\pmat}[1]{\begin{pmatrix}#1\end{pmatrix}}

\newcommand{\EE}{\mathsf{E}}
\newcommand{\PP}{\mathsf{P}}

\newcommand{\Let}{\coloneqq}

\DeclareMathOperator*{\minimize}{minimize}
\DeclareMathOperator{\sbjto}{subject\;to}

\begin{document}

\begin{frontmatter}

\title{Deterministic and probabilistic algorithms for stabilizing discrete-time switched linear systems\thanksref{footnoteinfo}}

\thanks[footnoteinfo]{This paper was not presented at any IFAC
meeting. Corresponding author Atreyee Kundu. Tel. +91-22-25764892.
Fax +91-22-25720057. N.\ Balachandran was supported in part by the grant 12IRCCSG016 from IRCC, IIT Bombay. D.\ Chatterjee was supported in part by the grant 12IRCCSG005 from IRCC, IIT Bombay.}

\author[Atreyee]{Atreyee Kundu}\ead{atreyee@sc.iitb.ac.in},
\author[Niranjan]{Niranjan Balachandran}\ead{niranj@math.iitb.ac.in},
\author[Atreyee]{Debasish Chatterjee}\ead{dchatter@iitb.ac.in}

\address[Atreyee]{Systems \& Control Engineering, Indian Institute of Technology Bombay, Powai, Mumbai - 400076, India}
\address[Niranjan]{Department of Mathematics, Indian Institute of Technology Bombay, Powai, Mumbai - 400076, India}




\begin{abstract}
     In this article we study algorithmic synthesis of the class of stabilizing switching signals for discrete-time switched linear systems proposed in \cite{knc_hscc14}. A weighted digraph is associated in a natural way to a switched system, and the switching signal is expressed as an infinite walk on this weighted digraph. We employ graph-theoretic tools and discuss different algorithms for designing walks whose corresponding switching signals satisfy the stabilizing switching conditions proposed in \cite{knc_hscc14}. We also address the issue of how likely/generic it is for a family of systems to admit stabilizing switching signals, and under mild assumptions give sufficient conditions for the same. Our solutions have both deterministic and probabilistic flavours.
\end{abstract}


\end{frontmatter}


\section{Introduction}
\label{s:intro}
    A \emph{switched system} comprises of two components --- a family of systems and a switching signal. The \emph{switching signal} selects an active subsystem from the family at every instant of time \cite[\S1.1.2]{Liberzon}, and the family of systems may contain nonlinear dynamics, systems with delays, etc. Switching signals are broadly classified as time-dependent (depends only on time), state-dependent (depends on state as well), and with memory (also depends on the history of active subsystems) \cite{Antsaklis_survey}.

    In this article we stay within the confines of discrete-time switched linear systems with purely time-dependent switching signals. Stability of such systems has been studied extensively by researchers over the past few decades, see e.g., \cite{Antsaklis_survey,Shorten_review,Heemels_survey,Liberzon} for detailed surveys. This study can be broadly classified into two categories --- stability under \emph{arbitrary switching} and stability under \emph{constrained switching}. In the former category, conditions on the family of systems are identified such that the resulting switched system is asymptotically stable under all admissible switching signals \cite{Daafouz02,Lee_Khar09,Lee_07,Lee005,Bliman_IFAC,Liberzon12}; in the latter category, given a family of systems, conditions on the switching signals are identified such that the resulting switched system is asymptotically stable \cite{knc_hscc14,Zhai002,Antsaklis_survey}.

    Our focus in this article is on stability of discrete-time switched linear systems under constrained switching \cite{knc_hscc14,abc}. Stability conditions in this direction primarily rely on the idea of \emph{slow switching} vis-a-vis \emph{(average) dwell time} switching. These results were originally developed in the context of continuous-time switched systems, but they can be readily extended to the discrete-time setting \cite{HespanhaMorse}, \cite[Chapter 3]{Liberzon}, with the (average) dwell time expressed in terms of the number of time steps \cite{Antsaklis_survey}. This classical theory pertains to switched systems in which all subsystems are stable. In case there are unstable systems, slow switching alone is not sufficient to guarantee stability of the switched system, and additional conditions are required to ensure that the switched system does not spend too much time on the unstable subsystems; see e.g., \cite{Antsaklis_survey}. Departing from this set of results, in \cite{Zhai002} the authors study global exponential stability of a discrete-time switched linear system in which not all subsystems are Schur stable (or possibly no subsystem is Schur stable) but the unstable subsystems form a stable combination. The characterization of the stabilizing switching signal in this new setting involves a modified definition of average dwell time \cite[Chapter 3]{Liberzon}, \cite{Zhai002}, and the rule of activating the Schur stable subsystems (if any) arbitrarily but activating the unstable subsystems depending on a prespecified ratio.

    In \cite{knc_hscc14} the authors present a general class of switching signals for global asymptotic stability of switched linear systems. Our conditions involve \emph{only} certain asymptotic properties of the switching signal and do not involve nor imply point-wise bounds on the number of switches --- unlike in the case of average dwell time switching. Consequently, there is plenty of flexibility insofar as the transient behaviour of the switching signals are concerned. Although this is not the first instance when unstable systems in the family are considered (e.g., \cite{Zhai002}), to the best of our knowledge, this is the first instance when unstable systems in the family are considered \emph{and} the stabilizing conditions depend only on asymptotic behaviour of the switching signal. On the one hand, our conditions require the presence of at least one asymptotically stable system in the family and is therefore ``conservative'' compared to the ones in the literature that accommodate families with all unstable systems (e.g., \cite{Zhai002}). On the other hand, we do not require the unstable systems in the family to form an asymptotically stable combination. The conditions proposed in \cite{knc_hscc14} are also numerically easier to verify than verifying semidefinite programming based conditions (e.g., \cite{Lee_Khar09}) on the given family of systems.

    In this article we address the important aspect of existence and algorithmic synthesis of the class of stabilizing switching signals proposed in \cite{knc_hscc14}.  Given a family of systems, the aspect of algorithmic synthesis of a class of stabilizing switching signals is important for obvious reasons and is not new in the switched systems literature, see e.g., \cite{Basar05}, where randomized algorithms to synthesize stabilizing state-dependent switching rules for multimodal systems are presented, and \cite{mitra_ADT}, where the authors propose optimization-based methods to verify the average dwell time property. In \cite{knc_hscc14} the authors provide a partial solution towards existence and algorithmic synthesis of the class of stabilizing switching signals under consideration. Here we extend the proposed techniques further and provide an array of results.

    The overarching contributions of this article are twofold, the first of which addresses the following pair of questions:
    \begin{enumerate}[label=(\alph*), leftmargin=*]
        \item Given a family of systems, possibly containing unstable dynamics, under what condition does there exist a switching signal that satisfies the conditions proposed in \cite[Theorem 1]{knc_hscc14}?
        \item If there exists such a switching signal, then how to detect/design it algorithmically?
    \end{enumerate}
    \begin{itemize}[label = $\circ$, leftmargin = *]
        \item Towards answering (a) and (b), we employ graph-theoretic terminology and arguments in our analysis. A weighted digraph is associated to a family of subsystems and the admissible transitions between them, and a switching signal is represented by an infinite walk on the above digraph, both in a natural way; see \S\ref{ss:swsigprop} for precise details. We are interested in the class of infinite walks corresponding to the class of stabilizing switching signals proposed in \cite{knc_hscc14}. We shall henceforth freely switch between system-theoretic and the corresponding graph-theoretic terminology in the above sense.
        \item We propose sufficient conditions for the desired infinite walk to admit what we call a closed contractive walk, the latter necessarily of finite length, and also discuss algorithmic detection/design of such walks. A closed contractive walk gives rise to an infinite walk corresponding to a stabilizing switching signal in a natural way, as we shall see momentarily.
        \item Given a weighted digraph, we provide two necessary and sufficient conditions for the existence of the aforementioned closed contractive walk in terms of a contractive circuit and a contractive cycle. For a contractive circuit we discuss an algorithm based on a linear program to detect such a circuit, and if it exists, we construct it algorithmically. For a contractive cycle we show that it is equivalent to detecting/designing a cycle of ``negative weight'' and employ standard algorithms for the same. See \S\ref{s:solA} for details.
    \end{itemize}
    The second major contribution of this article is towards posing and answering a more ambitious question:
    \begin{enumerate}[label=(\alph*), leftmargin = *, start=3]
        \item What class of switched systems admits the class of switching signals that satisfy the conditions proposed in \cite[Theorem 1]{knc_hscc14}?
    \end{enumerate}
    \begin{itemize}[label = $\circ$, leftmargin = *]
    	\item This question necessarily entails a shift in paradigm: while (a) and (b) dealt with conditions for a ``given'' switched system, (c) seeks to identify a suitable ``class of switched systems''. At the level of generality that (c) pertains to, there are two natural candidate apparatuses to turn to for providing its answers --- Baire category theory and probability. We turn to the latter, and under mild conditions provide a randomized algorithmic mechanism to identify a class of switched systems that satisfy the conditions proposed in \cite[Theorem 1]{knc_hscc14}.
    	\item Our contributions in this direction, presented in \S\ref{s:solB}, may be viewed from the following four perspectives:
       		\begin{itemize}[label = $\diamond$, leftmargin = *]
           		\item Firstly, our algorithm detects cycles on (the underlying digraph of) switched systems that are ``typical'' with high probability, and under mild conditions we guarantee that such cycles are contractive.
           	 	\item Secondly, the deterministic algorithms that are employed for answers to (a) and (b) above may not be applicable to switched systems whose underlying digraphs are large, especially if their sizes are so large that not all the weights can be kept in memory at once. For such large digraphs our algorithm provides probabilistic guarantees in the spirit of randomized algorithms for detection and design of contractive cycles.
            	\item Thirdly, and in cue with the preceding point, our algorithm is of an ``online'' nature in the following way: starting with a rough probabilistic description of the underlying weighted digraph, (i.e., without knowledge of the precise values of the weights,) we explore the digraph and synthesize a cycle during this exploration that is contractive with high probability. On the other hand, the traditional algorithms for detecting cycles require complete knowledge of the digraph and the vertex and edge weights a priori.
				\item Fourthly, if the constituent subsystems of a switched systems are prone to evolve over time in a manner that is not precisely known but certain statistical estimates of the nature of evolution are available, our algorithm applies and constructs a contractive cycle with uniform probabilistic guarantees over all such evolutions. See \S\ref{s:solB} for details.
	        \end{itemize}
    \end{itemize}

    The remainder of this article is organized as follows: In \S\ref{s:prelims} we briefly recall the class of stabilizing switching signals proposed in \cite{knc_hscc14} and formulate the problems under consideration. Our main results appear in \S\ref{s:solA} and \S\ref{s:solB}. We provide numerical examples in \S\ref{s:numex}. We conclude in \S\ref{s:concln} with a brief discussion on future directions. The proofs of our main results are provided in a consolidated fashion in \S\ref{s:proofs}.

    {\bf Notation}. $\N = \{1,2,\cdots\}$ is the set of natural numbers, $\N_{0} = \{0\}\cup\N$, and $\R$ is the set of real numbers. For a finite set $M$, $\abs{M}$ denotes its cardinality and $M\sqcup N$ denotes the disjoint union of $M$ with another finite set $N$. For a digraph $G(V,E)$ such that $V\Let V'\sqcup V''$, $d^{+}(v)$ denotes the outdegree of a vertex $v\in V$, $N^{+}_{V'} \Let \{u\in V'\:|\:(v,u)\in E\}$ denotes the set of outneighbours of a vertex $v$ in $V'$, and $d^{+}_{V'}(v) \Let \abs{N^{+}_{V'}}$ denotes the outdegree of $v$ in $V'$. For a walk $W$ on $G(V,E)$, $\abs{W}$ denotes its length.

\section{Preliminaries}
\label{s:prelims}
    We consider a family of discrete-time linear systems
    \begin{align}
    \label{e:family}
        x(t+1) = A_{i}x(t),\:\:x(0)\:\text {given},\:\:i\in\P,\:\:t\in\N_{0},
    \end{align}
    where $x(t)\in\R^{d}$ is the vector of states at time $t$, $\P = \{1,2,\cdots,N\}$ is a finite index set, and $A_{i}\in\R^{d\times d}$ is a known constant full-rank matrix for each $i\in\P$. Let $\sigma:\:\N_{0}\:\rightarrow\:\P$ be a \emph{switching signal} that specifies at every time $t$, the index of the active subsystem from the family \eqref{e:family}. The discrete-time \emph{switched linear system} generated by the given family of systems \eqref{e:family} and the switching signal $\sigma$ is given by
    \begin{align}
    \label{e:swsys}
        x(t+1) = A_{\sigma(t)}x(t),\:\:x(0)\:\text{given},\:\:t\in\N_{0}.
    \end{align}
    Given a family of systems \eqref{e:family}, in \cite{knc_hscc14} the authors identify a general class of switching signals $\sigma$ that admits a crisp characterization under which the resulting switched system \eqref{e:swsys} is globally asymptotically stable. Recall that by definition, the switched system \eqref{e:swsys} is \emph{globally asymptotically stable} (GAS) for a given switching signal $\sigma$ if \eqref{e:swsys} is
    \begin{itemize}[label=\(\circ\), leftmargin=*]
        \item Lyapunov stable, and
        \item globally asymptotically convergent: $\forall\,x(0)$, $\displaystyle{\lim_{t\rightarrow+\infty}x(t) = 0}$.
    \end{itemize}
    Preparatory to our results, we begin with:
    \subsection{Properties of the family \eqref{e:family}}
    \label{ss:familyprop}
    Let $\P_{S}$ and $\P_{U}\subset\P$ be the sets of indices of the asymptotically stable and unstable systems in the family \eqref{e:family}, respectively.
    \begin{fact}[{\cite[Fact 1]{knc_hscc14}}]
        \label{fact:key}
            For each $i\in\P$ there exists a pair $(P_{i},\lambda_{i})$, where $P_{i}\in\R^{d\times d}$ is a symmetric and positive definite matrix, and
        \begin{itemize}
            \item if $i\in\P_{S}$, then $0 < \lambda_{i} < 1$;
            \item if $i\in\P_{U}$, then $\lambda_{i} > 1$;
        \end{itemize}
        such that, with $\R^{d}\ni\xi\longmapsto V_{i}(\xi) \Let \xi^\top{P_{i}}\xi \geq 0$,
        we have $V_{i}(\gamma_{i}(t+1)) \leq \lambda_{i}V_{i}(\gamma_{i}(t)), t\in\N_{0}$, where $\gamma_{i}(\cdot)$ solves the $i-$th recursion in \eqref{e:family}, $i\in\P$. We call $V_{i}$, $i\in\P$, Lyapunov-like functions.
    \end{fact}
    \begin{fact}[{\cite[Fact 2, Proposition 1]{knc_hscc14}}]
    \label{fact:muij}
        There exist numbers $\mu_{ij}>0$ such that $V_{j}(\xi) \leq \mu_{ij}V_{i}(\xi)$ for all $\xi\in\R^{d}$, whenever it is admissible to switch from $i$ to $j$, $i,j\in\P$. In particular, the smallest such constants $\mu_{ij}$ are given by $\mu_{ij} = \lambda_{\max}(P_{j}P_{i}^{-1}),i,j\in\P$, where for a matrix $M\in\R^{n\times n}$ having real spectrum, $\lambda_{\max}(M)$ denotes its maximal eigenvalue.
    \end{fact}

    \subsection{Properties of the switching signal}
    \label{ss:swsigprop}

        We associate a weighted digraph $G(\P,E(\P))$ with a switched system in the following fashion:
        \begin{itemize}[label=$\circ$, leftmargin=*]
            \item The index set $\P$ denotes the set of vertices of $G$,
            \item The set of edges $E(\P)$ of $G$ consists of:
             \begin{itemize}[label=$\diamond$, leftmargin=*]
                \item a directed edge from $i$ to $j$, $i,j\in\P$, whenever a switching from system $i$ to system $j$ is admissible, and
                \item a self-loop at vertex $j$, $j\in\P$, whenever it is admissible to dwell on system $j$ for at least two consecutive time steps.
             \end{itemize}
             \item $w(i,j) \Let \ln\mu_{ij}$, $(i,j)\in E(\P)$ (\`a la Fact \ref{fact:muij}) and $w(j) \Let \abs{\ln\lambda_{j}}$, $j\in\P$, (\`a la Fact \ref{fact:key}) denote the edge weights and vertex weights of $G(\P,E(\P))$, respectively.
        \end{itemize}
        We may abbreviate $G(\P,E(\P))$ by $G$ if there is no risk of confusion. Recall that a \emph{walk} $W$ on a digraph $G(V,E)$ \cite[p.\ 4]{Bollobas} is an alternating sequence of vertices and edges $W = x_{0},e_{1},x_{1},e_{2},x_{2},\cdots,x_{\ell-1},e_{\ell},x_{\ell}$, where $x_{i}\in V$, $e_{i} = (x_{i-1},x_{i})\in E$, $0 < i \leq\ell$. The \emph{initial vertex} of $W$ is $x_{0}$ and the \emph{final vertex} of $W$ is $x_{\ell}$. If $x_{0} = x_{\ell}$, we say that the walk is closed. In this article we follow the convention: A closed walk $W = x_{0},(x_{0},x_{1}),x_{1},\cdots,x_{\ell-1},(x_{\ell-1},x_{0}),x_{0}$ is a \emph{circuit} if all its edges are distinct. $W$ is said to be a \emph{cycle} if the vertices $x_{i}$, $0 < i < \ell$ are distinct from each other and $x_{0}$. The length of a walk \cite[p.\ 5]{Bollobas} is its number of edges, counting repetitions, e.g., in the above case $W$ the length of the walk $W$ is $\ell$. In the sequel by the term \emph{infinite walk} we mean a walk of infinite length, i.e., it has infinitely many edges. An \emph{initial} \emph{subwalk} $W'$ of a walk $W$ is an initial segment of $W$, which we write as $W' \leq W$. The following essentially obvious fact associates a switching signal $\sigma$ to an infinite walk on the weighted digraph $G$:
        \begin{fact}[{\cite[Fact 3]{knc_hscc14}}]
        \label{fact:walk}
            The set of switching signals $\sigma:\:\N_{0}\to\P$ and the set of infinite walks on $G(\P,E(\P))$ (defined as above) are in bijective correspondence.
        \end{fact}
        For a walk $W$ on $G$:
        \begin{itemize}[label = $\circ$, leftmargin = *]
            \item Let $N_{W}$ be the number of distinct vertices that appear in $W$, and define
            \begin{align}
            \label{e:transfreq}
                \nu(W) \Let \frac{N_{W}}{\abs{W}},\quad \abs{W} > 0
            \end{align}
            to be the \emph{transition frequency} of $W$;
            \item we define the function
            \begin{align}
            \label{e:Xidefn}
                \Xi(W) \Let \frac{\displaystyle{\sum_{\mathrlap{(k,\ell)\in E(\P)}}w(k,\ell)\sharp\{k\rightarrow\ell\}_{W} + \sum_{\mathrlap{j\in\P_{U}}}w(j)\sharp\{j\}_{W}}}{\displaystyle{\sum_{\mathrlap{j\in\P_{U}}}w(j)\sharp\{j\}_{W}}},
            \end{align}
            where $\sharp\{k\rightarrow\ell\}_{W}$ and $\sharp\{j\}_{W}$ denote the number of times the edge $(k,\ell)$ and the vertex $j$ appear in $W$, respectively.
        \end{itemize}
        \subsection{The main result of \cite{knc_hscc14}}
        \label{ss:recallres}
        In the light of Fact \ref{fact:walk} we rephrase \cite[Theorem 1]{knc_hscc14} in the following manner:
        \begin{thm}[{\cite[Theorem 1]{knc_hscc14}}]
        \label{t:recallthm}
            Consider the underlying weighted digraph $G$ of the switched system \eqref{e:swsys}. The switched system \eqref{e:swsys} is GAS under all switching signals $\sigma$, whose corresponding infinite walks (\`a la Fact \ref{fact:walk}) $W$ satisfy
            \begin{align}
            \label{e:condn1}
                \varliminf_{\substack{\abs{W'}\rightarrow+\infty\\W'\leq W}}\nu(W') > 0,
            \end{align}
            and
            \begin{align}
            \label{e:condn2}
                \varlimsup_{\substack{\abs{W'}\rightarrow+\infty\\W'\leq W}}\Xi(W') < 1,
            \end{align}
            where $\nu(W')$ and $\Xi(W')$ are as defined in \eqref{e:transfreq} and \eqref{e:Xidefn}, respectively.
        \end{thm}
        \begin{rem}
    \label{r:condn1}
    \rm
        Since we are in the discrete-time setting, the association (\`a la Fact \ref{fact:walk}) of the length of a walk with time is natural. Condition \eqref{e:condn1} in the above theorem corresponds to the condition that the switching frequency of $\sigma$ is not asymptotically vanishingly small \cite[Theorem 1]{knc_hscc14}. In the presence of unstable systems in \eqref{e:family}, this condition is necessary to ensure that the switched system \eqref{e:swsys} does not eventually adhere to an unstable system. The first term in the numerator of $\Xi(W')$ captures the number of times each admissible transition $(k,\ell)\in E(\P)$ occurs in $\sigma$ till time $t$, weighted by $w(k,\ell) = \ln\mu_{k\ell}$, where $\mu_{k\ell}$ is as in Fact \ref{fact:muij}. The term $\displaystyle{\sum_{j\in\P_{S}}w(j)\sharp\{j\}_{W'}}$ (and $\displaystyle{\sum_{j\in\P_{U}}w(j)\sharp\{j\}_{W'}}$) captures the number of times system $j\in\P_{S}$ (resp. $\P_{U}$) is activated till time $\abs{W} = t$ by $\sigma$, weighted by $\abs{\ln\lambda_{j}}$, where $\lambda_{j}$ obeys Fact \ref{fact:key}.
    \end{rem}
    \begin{rem}
    \label{r:coselection}
    \rm
        Of course there is an element of ``choice'' in the selection of the Lyapunov-like functions in Fact \ref{fact:key} and consequently, $\lambda_{i}$'s and $\mu_{ij}$'s are not unique. Ideally one would like to algorithmically determine the possibility of co-designing the matrices $P_{i}$'s and the scalars $\lambda_{i}$'s such that switching signals satisfying \eqref{e:condn2} exist, and if so, to construct such a switching signal. This particular co-design problem, to our knowledge, is numerically difficult and in the absence of a numerical solution to it, we consider the matrices $P_{i}$'s and the scalars $\lambda_{i}$'s as given, and focus on algorithmic synthesis of switching signals satisfying \eqref{e:condn2}.
    \end{rem}

    Given a family of systems \eqref{e:family}, both
    \begin{enumerate}[label = (\alph*),leftmargin = *]
        \item the admissible transitions (connectivity of $G$), and
        \item the edge and vertex weights of $G$
    \end{enumerate}
    play a role in determining whether there exists an infinite walk $W$ that satisfies \eqref{e:condn2}. Indeed:
    \begin{description}[leftmargin = *]
        \item[Effect of (a)] Consider $\P = \{1,2,3\}$ with $A_{1} = \pmat{0.2 & 0.4\\0.6 & 0.1}$, $A_{2} = \pmat{0.1 & 0.9\\0.8 & 1.0}$ and $A_{3} = \pmat{1.0 & 0.3\\0.7 & 1.2}$. Consequently, $\P_{S} = \{1\}$ and $\P_{U} = \{2,3\}$. Let $E(\P) = \{(2,3),(3,2)\}$. In this case, even without the knowledge of the edge and vertex weights, we can conclude that there exists no infinite walk on $G$ that satisfies condition \eqref{e:condn2} because the term in the denominator of $\Xi(W')$, i.e., $\sharp\{1\}_{W'} = 0$ for all walks on $G$.
        \item[Effect of (b)] Consider the given family of systems \eqref{e:family} as in the above case. Assume that $E(\P) = \{(1,2),(2,1)\}$. From Fact \ref{fact:key} and Fact \ref{fact:muij}, we have the following estimates: $\mu_{12} = 0.8878$, $\mu_{21} = 1.7586$, $\lambda_{1} = 0.4314$ and $\lambda_{2} = 4.0281$. The term $w(1,2)\sharp\{1\rightarrow2\}_{W'} + w(2,1)\sharp\{2\rightarrow1\}_{W'} + w(2)\sharp\{2\}_{W'} = -0.1190\sharp\{1\rightarrow2\}_{W'} + 0.5645\sharp\{2\rightarrow1\}_{W'} + 1.3933\sharp\{2\}_{W'}$ is greater than the denominator $w(1)\sharp\{1\}_{W'} = 0.8407\sharp\{1\}_{W'}$ for all walks on $G$. Consequently, condition \eqref{e:condn2} is not satisfied.
    \end{description}

    In view of the above observations, given the underlying weighted digraph of the switched system \eqref{e:swsys}, we arrive at the important and natural question:
        \begin{enumerate}[label=\textsf{Problem \Alph*}, leftmargin=*, align=left, widest=B]
        	\item \label{problem:Aprime}\emph{Given a weighted digraph $G(\P,E(\P))$, does there exist an infinite walk \(W\) on $G(\P,E(\P))$ that satisfies \eqref{e:condn2}? If yes, can we provide a mechanism to detect/synthesize it?}
        \end{enumerate}
        In the language of switched systems, of course, \ref{problem:Aprime} is: ``Does there exist a switching signal $\sigma$ satisfying \eqref{e:condn2}? If such a $\sigma$ exists, then can we provide an algorithmic mechanism to detect/synthesize it?''

        \begin{rem}
        \label{r:infwalk}
        \rm
            It is important to clarify what we mean by algorithmic solutions to \ref{problem:Aprime}. We provide an algorithm that consists of (i) a finite walk $W_{0}$ of length $n_{0} > 0$, and (ii) an iterative process consisting of a mechanism requiring a bounded quantum of memory, to generate finite walks $W_{k}$ of length $n_{k} > 0$, \(k\in\N\), satisfying the condition that the final vertex of \(W_{k-1}\) is identical to the initial vertex of \(W_k\) for each $k$. We build the infinite walk \(W\) as the limit of $W_{1} W_2\cdots W_{k-1}W_k$, \(k\in\N\).
        \end{rem}
         Suppose that for a weighted digraph $G$ there is a self-loop at a vertex $j\in\P_{S}$. Consider a walk $W$ on $G$ such that it begins at this vertex $j$ and keeps on traversing that self-loop repeatedly. In this case $w(j,j) = 0$ by Fact \ref{fact:muij} and no vertex $j\in\P_{U}$ is visited. Consequently, the infinite walk generated satisfies \eqref{e:condn2}. Given a weighted digraph $G$, the above walk $W$ can be obtained from an algorithm that detects the above vertex $j\in\P_{S}$.

         However, given a weighted digraph $G$, detection of an infinite walk $W$ satisfying \eqref{e:condn2} is not simple beyond the above trivial case. Indeed, finding an infinite walk on $G$ that satisfies some prespecified condition involving the vertex and edge weights of $G$ is a computationally difficult problem. We define
        \begin{defn}
        \label{d:contrawalk}
            A walk $W$ on the weighted digraph $G(\P,E(\P))$ as \emph{contractive} if
            \begin{align}
            \label{e:contra}
                \Xi(W) < 1.
            \end{align}
        \end{defn}
        Given a weighted digraph $G$, we provide:
        \begin{enumerate}[label=\textsf{Solution \Alph*}, leftmargin=*, align=left, widest=B]
            \item \label{solution:Aprime} We establish a sufficient condition for the existence of an infinite walk satisfying \eqref{e:condn2} in terms of a closed contractive walk (necessarily of finite length) on $G$. This settles \ref{problem:Aprime}. We propose algorithmic techniques for synthesis of the above closed walk. Towards this, we derive a set of necessary and sufficient conditions in terms of a contractive \emph{circuit} and a contractive \emph{cycle} for the existence of a closed contractive walk on $G$, and apply numerically tractable algorithms to detect/design this circuit and/or cycle on $G$.
        \end{enumerate}
        Moving a step ahead from \ref{problem:Aprime}, and entailing a shift in paradigm, we pose:
        \begin{enumerate}[label=\textsf{Problem B}, leftmargin=*]
            \item \label{problem:Cprime}\emph{What class of weighted digraphs admits an infinite walk that satisfies \eqref{e:condn2}?}
        \end{enumerate}
        At the level of abstractness that \ref{problem:Cprime} pertains to, there are two natural apparatuses to turn to: the first is the Baire Category theorem and its consequences, and the second is probability theory. Considering the ensemble of switched systems as the sample space, we ask how likely is a switched system sampled from this ensemble to admit closed contractive switching signals, and to \ref{problem:Cprime} we provide:
        \begin{enumerate}[label = \textsf{Solution B}, leftmargin=*]
            \item \label{solution:Cprime} We propose a polynomial time algorithm that detects a \emph{cycle} of a certain fixed maximal length on $G$. Under mild assumptions on the connectivity and the weights associated to the vertices and edges of $G$ we provide probabilistic guarantee that the above cycle is contractive. Other perspectives and salient features of our algorithm have already been mentioned in the Introduction.
        \end{enumerate}
        \ref{solution:Aprime} and \ref{solution:Cprime} are provided in the following two sections \ref{s:solA} and \ref{s:solB}, respectively.

\section{\ref{solution:Aprime}}
\label{s:solA}
    In this section we expose \ref{solution:Aprime}. Since our solutions must be algorithmic (see Remark \ref{r:infwalk}), we specialize to finitary objects directly in:
    \begin{lem}[{\cite[Theorem 2(a)]{knc_hscc14}}]
    \label{lem:closedwalk}
        Consider the underlying weighted digraph $G(\P,E(\P))$ of the switched system \eqref{e:swsys}. If there exists a closed contractive walk $W$ on $G(\P,E(\P))$, then the infinite walk obtained by repeating the closed walk $W$ satisfies \eqref{e:condn2}.
    \end{lem}
     The task of algorithmic detection of a closed contractive walk on $G$ is computationally simpler under Lemma \ref{lem:closedwalk} since the length of the walk is finite.
    \begin{conv}
    \label{conv:weightconv}
        The total number of times a closed walk $W$ visits a vertex $j\in\P$ is the same as the total number of times $W$ visits the outgoing edges of the vertex $j\in\P$. Consequently, for a vertex $j\in\P$, $\sharp\{j\}_{W}$ can be replaced by $\displaystyle{\sum_{(j,\ell)\in E(\P)}\sharp\{j\rightarrow\ell\}_{W}}$. Since we are concerned with an infinite walk constructed by repeating the closed contractive walk $W$ indefinitely many times, the above convention is no loss of generality.
    \end{conv}
    Following Convention \ref{conv:weightconv}, the condition \eqref{e:contra} becomes
    \begin{align}
    \label{e:negativeclosed}
        \overline\Xi(W) \Let \sum_{(k,\ell)\in E(\P)}& \Bigl(w(k,\ell)+w(k){1}_{\P_{U}}(k)\nonumber\\
		&\;-w(k){1}_{\P_{S}}(k)\Bigr) \sharp\{k\rightarrow\ell\}_{W} < 0.
    \end{align}
    The mechanism explained in Remark \ref{r:infwalk} shows that for a walk $W$ generated by concatenating the walks $W_{1}$ and $W_{2}$ satisfying the usual contractivity condition, we have $\overline{\Xi}(W) = \overline{\Xi}(W_{1}) + \overline{\Xi}(W_{2})$.

    However, algorithmic detection of a closed contractive walk on $G$ is also difficult due to the absence of a bound on the length of the closed walk $W$. Consequently, the length at which the algorithm that attempts to detect a closed contractive walk should terminate must be specified and its selection is a difficult task a priori. A natural alternative is to specialize the closed walk $W$ to a walk of bounded length, for example, a circuit or a cycle. Our first main result provides necessary and sufficient conditions for the existence of a closed contractive walk on $G$ in terms of a contractive circuit and a contractive cycle:

    \begin{thm}
    \label{t:mainres1}
        Consider the underlying weighted digraph $G(\P,E(\P))$ of the switched system \eqref{e:swsys} as discussed in \S\ref{s:prelims}. The following are equivalent:\\\\
        i) $G(\P,E(\P))$ admits a closed contractive walk,\\
        ii) $G(\P,E(\P))$ admits a closed contractive circuit,\\
        iii) $G(\P,E(\P))$ admits a closed contractive cycle.\\\\
        Consequently, the infinite walk obtained by repeating one of the above satisfies \eqref{e:condn2}.
    \end{thm}
    See \S\ref{s:proofs} for a detailed proof of Theorem \ref{t:mainres1}.
    \begin{rem}
    \rm
        Theorem \ref{t:mainres1} gives a set of necessary and sufficient conditions for the existence of a closed contractive walk on a given weighted digraph $G$. We now seek algorithms that detect/design a contractive circuit or a contractive cycle on $G$. This task is numerically simpler compared to detection/design of a closed contractive walk of some length that is not known a priori for obvious reasons. In the remainder of this section we address these two algorithmic detection/design issues. An algorithm that detects a contractive circuit on the given weighted digraph $G$ is discussed in \cite[Theorem 2(b) and (c)]{knc_hscc14}. Here we improve upon the above algorithm to find a contractive circuit $W$ on $G$ that minimizes $\overline\Xi(W)$, if one such circuit $W$ exists. As regard to a contractive cycle, we demonstrate the applicability of the existing algorithms in our context.
    \end{rem}

    \subsection{Algorithm to detect/design contractive circuits}
    \label{ss:algocircuit}
    Given a weighted digraph $G$, our algorithm for detection/design of a contractive circuit on $G$ is motivated by the shortest path algorithm proposed in \cite[\S3.4]{papa_optimization}.
    \begin{prop}[{\cite[Theorem 2(b),(c)]{knc_hscc14}}]
        \label{prop:algo}
            Let the underlying weighted digraph $G(\P,E(\P))$ of the switched system \eqref{e:swsys} as defined in \S\ref{s:prelims} be given.
            \begin{description}
            \item[Step 1]A contractive circuit $W$ on $G(\P,E(\P))$ that minimizes $\overline\Xi(W)$ is obtained from the solution to the following feasibility problem in the variable $\eta\in\R^{\abs{E(\P)}}$:
            \begin{align}
            \label{e:feasprob}
                \minimize_W &\quad \overline\Xi(W) \\
                \sbjto &\quad
                \begin{cases}
                    A\eta = 0\in\R^{\abs{\P}},& \nonumber\\
                    \overline\Xi(W) < 0, &\nonumber\\
                    0 \leq \eta_{j} \leq 1\:\:\text{for all}\:\: j = 1,2,\cdots,\abs{E(\P)}, &\nonumber\\
                    \displaystyle{\sum_{j=1}^{\abs{E(\P)}}\eta_{j}} \geq 1,& \nonumber
                \end{cases}
            \end{align}
            where $A$ is the node (arc) incidence matrix of $G(\P,E(\P))$.\footnote{The node (arc) incidence matrix $A = [a_{ij}]$ \cite[\S3.4]{papa_optimization} of $G$ is defined by%
        \begin{align}
        \label{e:incimat}
            a_{ij} =
            \begin{cases}
                +1	& \text{if edge $(i,j)$ leaves node $i,\:\:i = 1,\cdots,\abs{\P}$},\\
                -1	& \text{if edge $(i,j)$ enters node $i,\:\:j = 1,\cdots,\abs{E(\P)}$},\\
                0	& \text{otherwise}.
            \end{cases}
        \end{align}}
            \item[Step 2]From the solution to the feasibility problem \eqref{e:feasprob}, a contractive circuit $W$ on $G(\P,E(\P))$ that minimizes $\overline\Xi(W)$ can be obtained by the application of Hierholzer's algorithm.\footnote{See \cite[p. 57]{Harris} for Hierholzer's algorithm.}
        \end{description}
        \end{prop}
        \begin{rem}
        \rm
        a) There are two steps to the algorithm in Proposition \ref{prop:algo}: In Step 1, we employ a feasibility problem \eqref{e:feasprob} to determine a contractive circuit $W$ on $G(\P,E(\P))$ that minimizes $\overline\Xi(W)$, if $G(\P,E(\P))$ admits such a circuit. The feasibility problem \eqref{e:feasprob} involves solving a linear program for the vector $\eta$. Even though \eqref{e:feasprob} is a bona fide linear program, \cite[Corollary to Theorem 13.3]{papa_optimization} guarantees that it has integer optimal solutions. In other words, the condition $0 \leq \eta_{j} \leq 1$ for all $1 \leq j \leq \abs{E(\P)}$ implies that each entry of the feasible solution vector $\eta$ is either $1$ or $0$, corresponding to whether an edge is included in the circuit or not, respectively. By definition of a circuit, the size of the vector $\eta$ is at most $\abs{E(\P)}$. If the feasibility problem \eqref{e:feasprob} has a solution, we obtain a subgraph on $G$ from the vector $\eta$ that admits a circuit satisfying condition \eqref{e:negativeclosed}. If the feasibility problem \eqref{e:feasprob} has a solution, we proceed to Step 2 and apply Hierholzer's algorithm to find such a circuit on $G$. Hierholzer's algorithm admits the above subgraph as input in Step 2. Recall \cite[p.\ 57]{Harris} that given an Eulerian graph $\mathcal{G}$, Hierholzer's algorithm finds an Eulerian circuit of $\mathcal{G}$. The applicability of this algorithm in our context is explained in detail in the proof of Proposition \ref{prop:algo} \cite[\S6.2]{knc_hscc14}.\\
        b) The condition $A\eta = 0\in\R^{\abs{\P}}$ in the feasibility problem \eqref{e:feasprob} corresponds a circuit, and this equation always has a trivial solution where vector $\eta$ has all entries equal to $0$. The condition $\displaystyle{\sum_{j=1}^{\abs{E(\P)}}\eta_{j} \geq 1}$ prevents the above trivial solution.
        \end{rem}
        \begin{rem}
        \label{r:algomod}
        \rm
            With respect to \cite[Theorem 2(a) and (b)]{knc_hscc14}, we have modified the ``objective function'' in \eqref{e:feasprob} above. This ensures that the solution to the feasibility problem \eqref{e:feasprob} (if any) corresponds to a contractive circuit on $G$ that minimizes $\overline\Xi(W)$.
        \end{rem}

        \subsection{Algorithms to detect/design contractive cycles}
        \label{ss:algocycle}
         Given a weighted digraph $G$, the algorithmic detection/design of a contractive cycle is equivalent to finding what is commonly known as a negative cycle. Indeed, a negative cycle is one for which the sum of the edge weights is less than zero, which is precisely condition \eqref{e:negativeclosed}. A large class of algorithms is available to achieve the above; see e.g., \cite{Lewandowski10,Zaroliagis08} for detailed surveys. Perhaps the most well-known one in this class is the Bellman-Ford-Moore algorithm, which is a shortest path algorithm, 
		 and it detects and reports negative weight cycles that are reachable from a pre-specified source vertex \cite[p. 646]{Cormen_algo}. In our context, the Bellman-Ford-Moore algorithm suffices insofar as the detection of a contractive cycle on $G$ is concerned; further work is however required for constructing a negative cycle if such a cycle exists. Beyond the Bellman-Ford-Moore algorithm, a newer algorithm proposed in \cite{allnegcycles} lists all elementary negative cycles on a given weighted digraph; this particular algorithm jointly serves purposes of detection and design of negative cycles.

        \begin{rem}
        \label{r:circuitvscycle}
        \rm
            Since we are interested in detecting/designing \emph{an} infinite walk that satisfies \eqref{e:condn2}, algorithmically finding either a contractive circuit or a contractive cycle and designing an infinite walk by repeating the above suffices. However, the question of whether to find a contractive circuit or a contractive cycle algorithmically based on the following aspect appears to be interesting: Consider minimizing $\overline\Xi(W)$ in the sense of maximally negative $\overline\Xi(W)$ over all contractive circuits or cycles $W$ on a given weighted digraph $G$.
            \begin{itemize}[label = $\circ$, leftmargin = *]
                \item On the one hand, a contractive cycle is necessarily a contractive circuit.\footnote{If an edge appears more than once in a walk, the corresponding vertices are also repeated.} On the other hand, although a contractive circuit is not necessarily a contractive cycle, it necessarily contains at least one contractive cycle.\footnote{This is evident from the proof of Theorem \ref{t:mainres1}.} Consequently, a contractive circuit $W$ on $G$ that minimizes $\overline\Xi(W)$ may contain one or more contractive cycles.
                \item The algorithm in Proposition \ref{prop:algo} detects a contractive circuit $W$ on $G$ that minimizes $\overline\Xi(W)$. If this circuit is only a contractive cycle, then the output of the above algorithm is a contractive cycle as well.
                \item To obtain a cycle $W$ on $G$ that minimizes $\overline\Xi(W)$ by the application of negative cycle detection/design algorithms, a two step mechanism is required: first, to list all elementary negative cycles on $G$ (by applying the algorithm in \cite{allnegcycles}) and second, to find the most negative one from this list.
                \item Instead of applying the above mechanism to $G$, we may apply it to the subgraph of $G$ obtained from the feasibility problem \eqref{e:feasprob} in Proposition \ref{prop:algo}. Since this subgraph admits a circuit $W$ that minimizes $\overline\Xi(W)$, it necessarily admits the most negative cycle on $G$. It is immediate that considering a subgraph of $G$ instead of $G$ reduces both time and space complexity associated to the search of the most negative cycle.
            \end{itemize}
        \end{rem}

    \section{\ref{solution:Cprime}}
    \label{s:solB}
    We first propose the following algorithm for detection of cycles in $\P_{S}$; it will be utilized in Theorem \ref{t:mainres2} below to furnish certain genericity assertions (see Remark \ref{r:mainres2}).
        \begin{alg}
        \label{algo:cyclealgo}
        \mbox{}
        \begin{description}
            \item[Step 1] Set $k = 0$.\\
                        Pick $j_{k}\in\P_{S}$ uniformly at random.
            \item[Step 2] If $N_{\P_{S}}^{+}(j_{k})\backslash\{j_{0},\cdots,j_{k}\} \neq \emptyset$,\\
                         \hspace*{0.5cm}Pick $j_{k+1}\in N_{\P_{S}}^{+}(j_{k})\backslash\{j_{0},\cdots,j_{k}\}$ uniformly at random.\\
                         \hspace*{0.5cm}Set $k = k + 1$.\\
                         \hspace*{0.5cm}Go to Step 2.\\
                         Else\\
                         \hspace*{0.5cm}Pick $j_{k+1} = j_{i}$ such that $j_{i}\in N_{\P_{S}}^{+}(j_{k})$ and $(k-i)$ is maximum.\\
                         \hspace*{0.5cm}Go to Step 3.
            \item[Step 3] End.
        \end{description}
        \end{alg}
        For example, let $\P = \{1,2,3,4,5\}$ with $\P_{S} = \{1,2,3\}$ and $\P_{U} = \{4,5\}$. Let
            \begin{align*}
             E(\P) = \{&(1,2),(1,3),(1,4),(1,5),
                (2,1),(2,3),\\&(2,4),(2,5),
                (3,2),(3,4),(3,5),
                (4,1),\\&(4,2),(4,3),(4,5)\}.
            \end{align*}
            Let $j_{0} = 1\in\P_{S}$. Then $j_{1}$ is selected from $\{2,3\}\backslash\{1\}$. Let $j_{1} = 2$. Then $j_{2}$ is selected from $\{1,3\}\backslash\{1,2\}$. Consequently, $j_{2} = 3$. Now, $\{2\}\backslash\{1,2,3\} = \emptyset$. As a result, $j_{3} = 2$. So we obtain the walk $1,(1,2),2,(2,3),3,(3,2),2$, which contains the cycle $2,(2,3),3,(3,2),2$.

            In the remainder of this section we show that under mild connectivity and generic weight assumptions on the given weighted digraph $G$, a cycle obtained from Algorithm \ref{algo:cyclealgo} satisfies \eqref{e:negativeclosed} with high probability.
         Let $\Phi:\N\rightarrow\R$ be a monotone increasing function.
        \begin{defn}
        \label{d:niceconnwt}
             A weighted digraph $G(\P,E(\P))$ is said to be
			 \begin{itemize}[label=\(\circ\), leftmargin=*, nolistsep]
			 	\item \emph{nicely connected} if $d_{\P_{S}}^{+}(j) \geq \lfloor\Phi(\abs{\P_{S}})\rfloor$ for all $j\in\P$;
				\item \emph{nicely weighted} if the vertex and edge weights on $G$ satisfy the following conditions:
		            \begin{itemize}[label=$\triangleright$, leftmargin = *, nolistsep]
       		        	\item there exist \(\beta, B > 0\) satisfying $0 < \beta < B$ such that the vertex weights $w(j)$ are independent and $0 < w(j) \leq B$ with $\EE[w(j)] = \beta$ for all $j\in\P$, and%
               			\item there exist constants $A > 0$ and $\alpha < \beta$ such that for every $(i,j)\in E(\P)$, the edge weight $w(i,j)\in[-A,A]$ and $\EE[w(i,j)]\leq\alpha$.
            		\end{itemize}
			\end{itemize}
        \end{defn}
        \begin{rem}
        \label{r:constancy of weight bounds}
        \rm
			The condition that the vertex and edge weights \(w(j)\) and \(w(i, j)\) are uniformly bounded if \(G\) is nicely weighted is no loss of generality on account of the graph \(G\) being finite. However, it is also possible to consider the case in which the bounds on the weights depend on the size of the graph \(G\), as explained in Remark \ref{r:variable weight bounds} below. We stick to the simpler case for ease of exposition.
        \end{rem}
    The following lemma guarantees the existence of a cycle in $\P_{S}$ of length at least $\lfloor\Phi(\abs{\P_{S}})\rfloor$.
        \begin{lem}
        \label{lem:cyclelem}
            If the given weighted digraph $G(\P,E(\P))$ is nicely connected, then Algorithm \ref{algo:cyclealgo} detects a cycle $W$ on $G(\P,E(\P))$ such that all vertices in $W$ are from $\P_{S}$ and the length of $W$ is at least $\lfloor\Phi(\abs{\P_{S}})\rfloor$.
        \end{lem}
        See \S\ref{s:proofs} for a short proof of Lemma \ref{lem:cyclelem}.
    Our final result is the following:
    \begin{thm}
        \label{t:mainres2}
            Consider the switched system \eqref{e:swsys} and the underlying weighted digraph $G(\P,E(\P))$ as described in \S\ref{s:prelims}. Suppose that $G(\P,E(\P))$ is nicely connected and nicely weighted. Then a cycle of length at least $\lfloor\Phi(\abs{\P_{S}})\rfloor$ on $G(\P,E(\P))$ obtained from Algorithm \ref{algo:cyclealgo} is contractive with probability at least
			\[
				1-\exp\biggl(-\frac{1}{2}\Bigl(\frac{\alpha-\beta}{A+B}\Bigr)^{2}\lfloor\Phi(\abs{\P_{S}})\rfloor\biggr).
			\]
            Consequently, the infinite walk obtained by repeating the above cycle satisfies \eqref{e:condn2}.
        \end{thm}
        We present a proof of Theorem \ref{t:mainres2} in \S\ref{s:proofs}.


        \begin{rem}
        \label{r:mainres2}
        \rm
             Theorem \ref{t:mainres2} asserts that a cycle obtained via Algorithm \ref{algo:cyclealgo} is contractive with high probability provided $\abs{\P_{S}}$ is large. Consequently, repeating such a cycle derived from Algorithm \ref{algo:cyclealgo} generates an infinite walk $W$ that, in view of Lemma \ref{lem:closedwalk}, satisfies \eqref{e:condn2}. This in turn identifies a class of switched systems (whose underlying weighted directed graph $G$ is nicely connected and nicely weighted) that admits switching signals satisfying the conditions proposed in \cite[Theorem 1]{knc_hscc14} with overwhelming probability.
        \end{rem}

        \begin{rem}
        \label{r:variable weight bounds}
        \rm
			The primary engine leading to the estimate in Theorem \ref{t:mainres2} is Azuma's inequality. Our assumption of a uniform bound for the weights due to \(G\) being nicely weighted led to a uniform bound on the martingale increments \((M_m - M_{m-1})_{m=1}^n\) in the proof of Theorem \ref{t:mainres2}, and our estimate followed at once from Azuma's inequality. A more general version of Azuma's inequality may be employed in an identical fashion to cater to vertex- and edge-dependent weights, leading to a possibly sharper bound. The numerical value of the confidence with which a contractive cycle may be found, however, depends on the size of \(\P_S\) and the ability of the function \(\Phi\) in Definition \ref{d:niceconnwt} to dominate the accumulation of the weights along the martingale increments.
        \end{rem}

        \begin{rem}
        \label{r:algofeatures}
        \rm
            Given the underlying weighted digraph $G$ of the switched system \eqref{e:swsys}, the deterministic algorithms for detection/design of a contractive cycle, as discussed in \S\ref{ss:algocycle}, require the complete knowledge of all the vertex and edge weights of $G$ prior to their application. Consequently, these are ill suited for large graphs where all the weights cannot be stored in the memory at once. In contrast, Algorithm \ref{algo:cyclealgo} explores $G$ without prior knowledge of the vertex and edge weights and during this exploration designs a cycle, which is contractive with high probability. In particular, this lends an ``online'' flavour to our algorithm. Consequantly, it is suited for a class of large weighted digraphs for which deterministic guarantees are difficult or impossible to give. In addition, consider the case when certain parameters of the subsystems in the given family \eqref{e:family} (and consequently the vertex and edge weights of $G$) evolve over time in a manner that is not completely known. A cycle obtained from Algorithm \ref{algo:cyclealgo} is contractive with high probability independent of this evolution as long as $G$ is nicely weighted.
        \end{rem}


\section{Numerical Examples}
\label{s:numex}
    \begin{exmp}
    \rm
        This example corresponds to our \ref{solution:Aprime}. Based on our discussion regarding ``contractivity'' in \S\ref{s:solA}, we select the algorithm in Proposition \ref{prop:algo} to find a ``most'' contractive circuit. We consider the family of systems \eqref{e:family} with $\P = \{1,2,3,4\}$, where
        \begin{align*}
            A_{1} &= \pmat{0.2 & -0.7\\0.8 & 0.3}, & A_{2} &= \pmat{0.5 & 0.1\\0.4 & 0.2},\\
            A_{3} &= \pmat{1.2 & 0.9\\1.4 & 0.2}, & A_{4} &= \pmat{1.1 & 0.2\\0.2 & 0.7}.
        \end{align*}
        For this family $\P_{S} = \{1,2\}$ and $\P_{U} = \{3,4\}$. Let all transitions among the systems in the given family be admissible. Let it also be permissible for switching signals to dwell on systems $3$ and $4$ for two (or more) consecutive time steps. That is,
        \begin{align*}
            E(\P) = \{&(1,2),(1,3),(1,4),
            (2,1), (2,3), (2,4),\\
            &(3,1),(3,2),(3,3),(3,4),
            (4,1),(4,2),\\&(4,3),(4,4)\}.
        \end{align*}
        We construct the node(arc) incidence matrix $A$ as described in \eqref{e:incimat}.\footnote{Incidence matrices are in general defined for graphs without self-loops. We accommodate self-loops in an incidence matrix in the following manner: to a vertex $j\in\P$ such that $j$ has a self-loop, we associate an auxiliary vertex $j'$. The transitions $j$ to $j'$ and $j'$ to $j$ represent the self-loops.} We avoid presenting the matrix here for reasons of space. The elements of the column vector $\eta$ are associated with the entries of $E(\P)$.

        For the given family of systems \eqref{e:family}, we obtain an estimate for the numbers $\lambda_{i}$ and $\mu_{ij}$ from Fact \ref{fact:key} and Fact \ref{fact:muij}, respectively.
        \begin{align*}
            \lambda_{1} &= 0.6480, & \lambda_{2} &= 0.4200, & \lambda_{3} &= 4.9946,\\
            \lambda_{4} &= 3.3657, & \mu_{12} &= 0.6094, & \mu_{13} &= 0.4067,\\
            \mu_{14} &= 0.4067, & \mu_{21} &= 2.4470, & \mu_{13} &= 0.9914,\\
            \mu_{14} &= 0.9914, & \mu_{31} &= 2.8406, & \mu_{32} &= 1.7241,\\
             \mu_{33} &= 1, & \mu_{34} &= 1, & \mu_{41} &= 2.8406,\\
              \mu_{42} &= 1.7241, & \mu_{43} &= 1, & \mu_{44} &= 1,
        \end{align*}
        and associate $\abs{\ln\lambda_{j}}$ and $\ln\mu_{ij}$ as vertex weights $w(j)$ and edge weights $w(i,j)$ of $G$, respectively.

        Solving the feasibility problem \eqref{e:feasprob} in the context of this setting with the aid of MATLAB by employing the program YALMIP \cite{Lofberg04} and the solver SDPT3 \cite{SDPT3}, we obtain the following solution:
        \[
            \eta = (1,0,0,1,0,0,0,0,0,0,0,0,0,0,0,0)^\top,
        \]
        with the corresponding $\overline\Xi(W) = -0.084851 < 0$. A circuit $W$ obtained from the vector $\eta$ with the aid of Hierholzer's algorithm is:
        \[
            1,(1,2),2,(2,1),1.
        \]
        We consider the switching signal corresponding to the infinite walk obtained by repeating the above circuit $W$ and study $(\norm{x(t)})_{t\in\N_{0}}$ for 100 different initial conditions chosen uniformly at random from the interval $[-1000,1000]^{2}$.
             \begin{figure}[htbp]
            \begin{center}
                \includegraphics[scale=0.45]{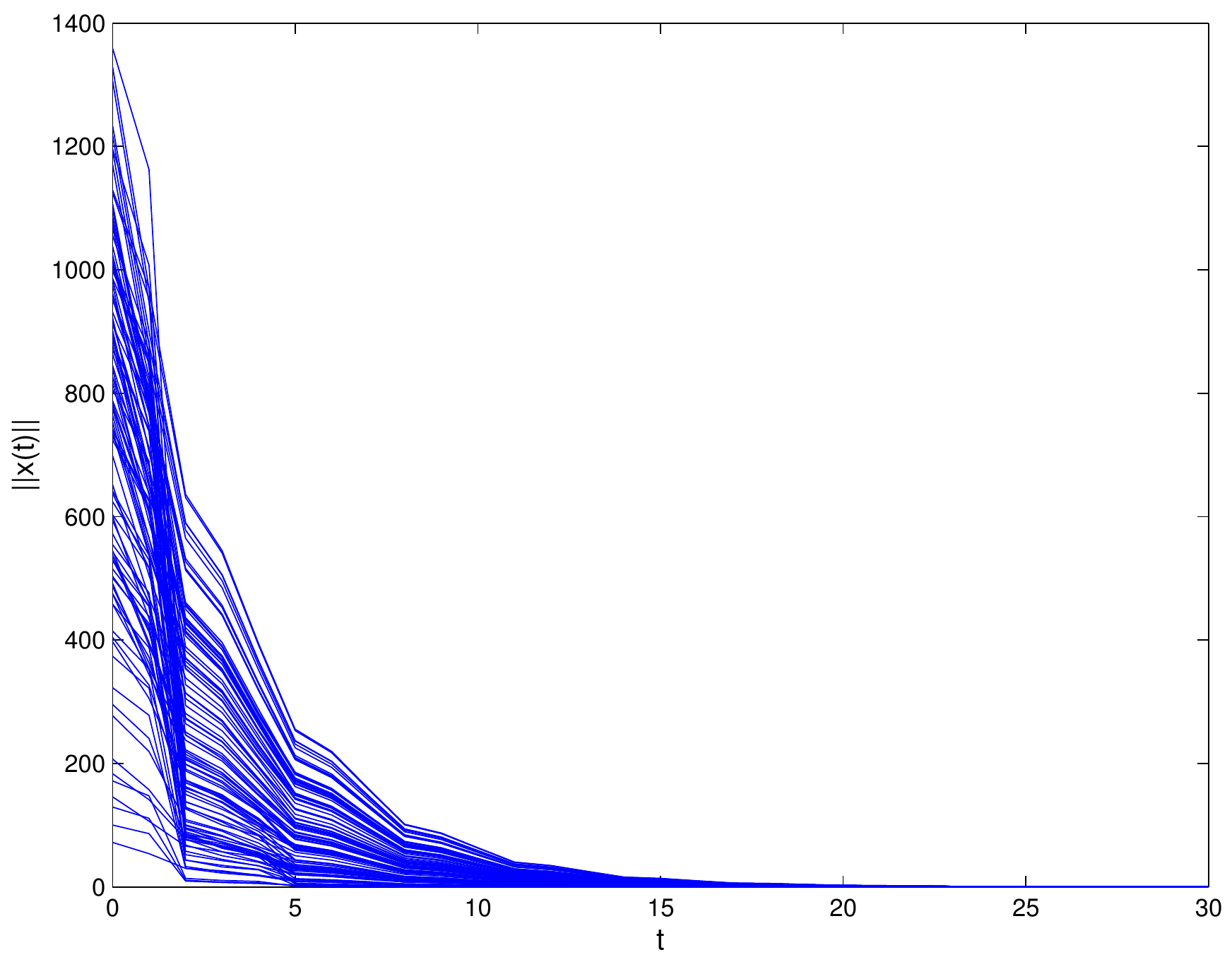}
            \end{center}
            \caption{Plot for $(\norm{x(t)})_{t\in\N_{0}}$ for 100 different initial conditions chosen uniformly at random from the interval $[-1000,1000]$.} \label{fig:swsig}
            \end{figure}
    \end{exmp}

    \begin{exmp}
    \rm
        This example corresponds to \ref{solution:Cprime}.
        Consider a nicely connected and nicely weighted directed graph $G$ with
            \begin{itemize}[label = $\circ$,leftmargin = *]
                \item $\abs{\P_{S}} = 1000$,
                \item $\Phi(r) = \frac{1}{10}\sqrt{r}$,
                \item $d^{+}(j) = \lfloor\Phi(\abs{\P_{S}})\rfloor$ for all $j\in\P$, and
                \item $A = 2.5$, $B = 5$, $\alpha = 0$ and $\beta = 2.5$.
            \end{itemize}
            We extract and fix a cycle $W$ obtained from Algorithm \ref{algo:cyclealgo} on $\P_{S}\subset\P$. The vertex and edge weights on $W$ are sampled uniformly at random $1000$ times from the intervals as stipulated in Definition \ref{d:niceconnwt}. We calculate $X_{n}$, as defined in \eqref{e:Xndefn} empirically for $n$ being the length of the cycle $W$.

            The above experiment is repeated for cycles of different length $n$ obtained from Algorithm \ref{algo:cyclealgo} with uniformly randomly selected initial vertex. We plot the empirical probability of $\{X_{n} < 0\}$ vs length $n$ of the cycle in Figure \ref{fig:plot}.
            \begin{figure}[htbp]
            \begin{center}
                \includegraphics[scale=0.45]{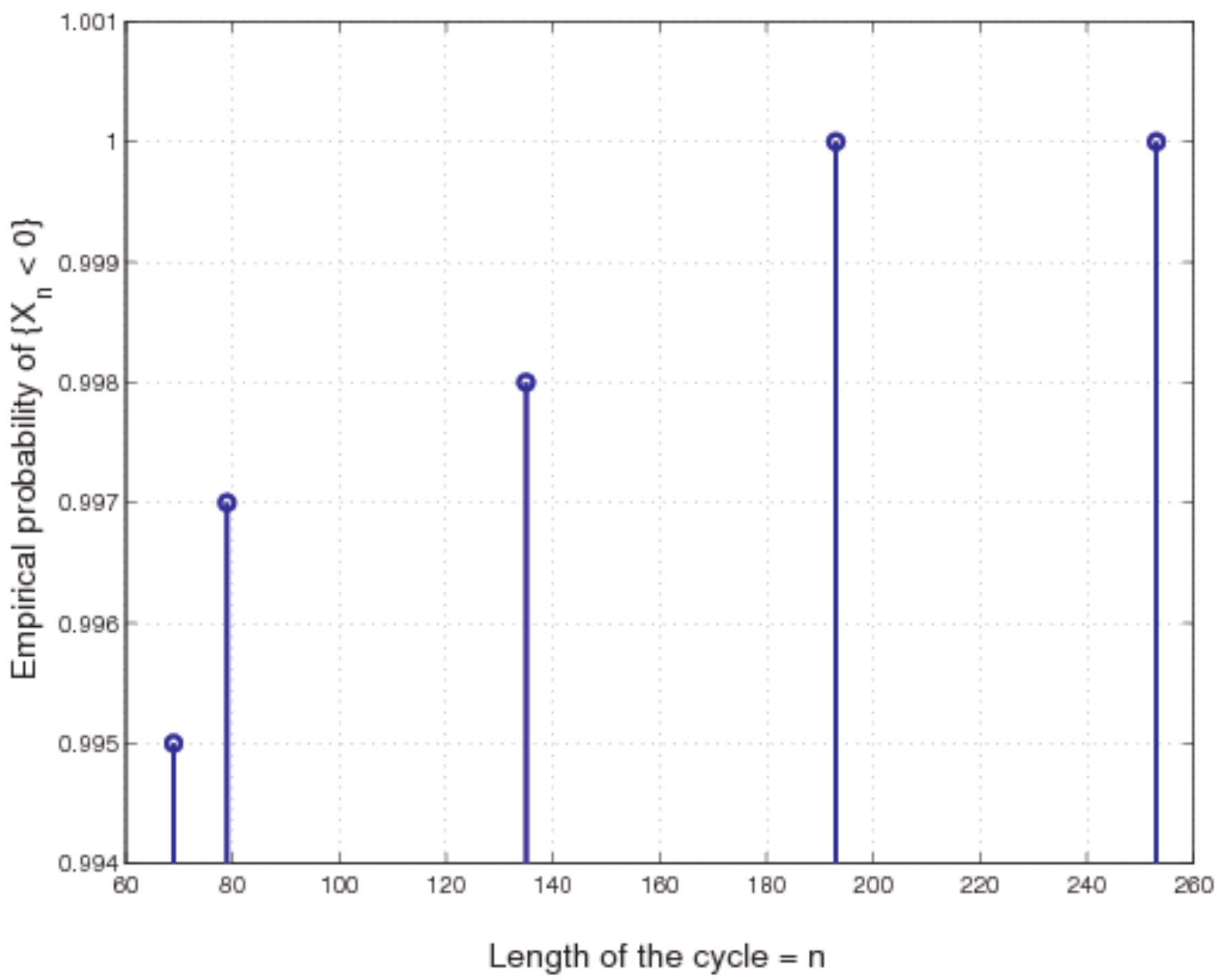}
            \end{center}
            \caption{Plot for empirical probability of the cycle being contractive vs the length of the cycle $n$ with $\displaystyle{\Phi(r) = \frac{1}{10}\sqrt{r}}$.} \label{fig:plot}
            \end{figure}
            Observe that the detection of a contractive cycle from Algorithm \ref{algo:cyclealgo} does not require a priori knowledge of the vertex and edge weights of $G$. It is evident from this example as we first fix a cycle $W$ and then select weights from a specified interval. This is not the case with deterministic negative cycle detection algorithms, which require complete knowledge of the vertex and edge weights of $G$ prior to their application. In addition, the weights are sampled uniformly at random 1000 times and we find high empirical probability for $\{X_{n} < 0\}$. This highlights the feature that even if the systems in the given family are prone to evolve over time, our algorithm provides uniform probabilistic guarantees.
    \end{exmp}
%
\section{Conclusion}
\label{s:concln}
    In this article we discussed several methods to algorithmically synthesize members of the class of stabilizing switching signals proposed in \cite{knc_hscc14}. A weighted digraph was associated to a switched system and the switching signal is expressed as an infinite walk on this weighted digraph in a natural way. In this setting we proposed a sufficient condition for the existence of an infinite walk whose corresponding switching signal satisfies the conditions in \cite[Theorem 1]{knc_hscc14}, and discussed algorithms for the synthesis of the above infinite walk. Thereafter we tackled the question of how likely is it for a ``generic'' switched system to admit a stabilizing switching signal, and we identified such a class of switched systems in terms of connectivity and vertex and edge weights of the underlying weighted digraph of the switched system. Necessary conditions for the existence of the above infinite walk is currently under investigation and will be reported elsewhere.

\section{Proofs of Main Results}
\label{s:proofs}
    {\bf Proof of Theorem \ref{t:mainres1}}: i) $\Rightarrow$ ii) Assume that the given weighted digraph $G$ admits a closed contractive walk $W = x_{0},(x_{0},x_{1}),x_{1},\cdots,x_{\ell-1},(x_{\ell-1},x_{0}),x_{0}$ of length $\ell$, but does not admit a contractive circuit. By assumption, $W$ is not a circuit. We claim that $W$ can be recursively decomposed into circuits.\\
    Suppose that $W$ has $n$ edges, $n\geq 1$, which appear more than once in $W$.\\
    \emph{Basis Step}: Pick any one of the edges, say $(x_{k-1},x_{k})$, that appears more than once in $W$. Obtain $W_{1}$ and $W_{2}$ such that $W_{1}$ is the subwalk generated by concatenating the following:
    \begin{itemize}[label = $\circ$, leftmargin = *]
        \item $x_{0},(x_{0},x_{1}),x_{1},\cdots,x_{k-1},(x_{k-1},x_{k})$, i.e., the walk from the beginning of $W$ till the \emph{first} instance of $(x_{k-1},x_{k})$ in $W$, and
        \item $x_{k},(x_{k},x_{k+1}),x_{k+1},\cdots,x_{\ell-1},(x_{\ell-1},x_{0}),x_{0}$, i.e., the walk beginning after the \emph{last} instance of $(x_{k-1},x_{k})$ in $W$ till the end of $W$;
    \end{itemize}
    and $W_{2} = x_{k},\cdots,x_{k}$ is the subwalk generated by removing $W_{1}$ from $W$. Clearly, $W_{1}$ is a closed walk with at most $(n-1)$ edges, which appear more than once, and $W_{2}$ is a closed walk with at most $n$ edges, that appear more than once. In case $n=1$, $W_{1}$ is a circuit. If in addition $(x_{k-1},x_{k})$ appears exactly twice, $W_{2}$ is also a circuit. Since $W$ satisfies $\overline\Xi(W) < 0$, one of the following three conditions holds: i. $\overline\Xi(W_{1}) < 0$, ii. $\overline\Xi(W_{2}) < 0$, iii. both $\overline\Xi(W_{1})$ and $\overline\Xi(W_{2}) < 0$.\\
    \emph{Recursive Step}: In case one of the first two conditions holds, we select the subwalk $W_{i}$ that satisfies $\overline\Xi(W_{i}) < 0$. In case the last condition holds, we select the subwalk $W_{i}$ with fewer number of edges, that appear more than once. If the selected $W_{i}$ is a circuit, we stop; else we generate $W_{i}^{(1)}$ and $W_{i}^{(2)}$ by the same procedure explained in Basis Step and continue till we obtain $W_{i}^{(j)}$, $j\in\{1,2\}$ as a circuit.

    Armed with the above claim, consider the case when a closed contractive walk $W$ that satisfies $\overline\Xi(W) < 0$ is decomposed into $p$ circuits --- $W_{1},W_{2},\cdots,W_{p}$. Now, since $\overline\Xi(W) < 0$, at least one of the $W_{i}$'s, $i\in\{1,2,\cdots,p\}$, satisfies $\overline\Xi(W_{i}) < 0$. A circuit $W_{i}$ satisfying $\overline\Xi(W_{i}) < 0$ is the one that we are looking for. It contradicts our assumption that there is no circuit on $G$ that is contractive.

    ii) $\Rightarrow$ iii) Assume that there is a contractive circuit $W = x_{0},(x_{0},x_{1}),x_{1},\cdots,x_{\ell-1},(x_{\ell-1},x_{0}),x_{0}$ of length $\ell$ on the given weighted digraph $G$ but no contractive cycle. By assumption, $W$ is not a cycle. We claim that $W$ can be recursively decomposed into cycles.\\
    \emph{Basis Step}: \textsf{Step 1}. Suppose $x_{0}$ appears $n$ times other than at the first and last positions of $W$. If $n=0$, apply Step 2 on $W$; else decompose $W$ into subwalks $W_{1},W_{2},\cdots,W_{n+1}$ in the following fashion:
    \begin{itemize}[label= $\circ$, leftmargin = *]
        \item $W_{1}$ is the subwalk from the beginning of $W$ till the first repeated instance of $x_{0}$,
        \item $W_{2}$ is the subwalk beginning from the first repeated instance of $x_{0}$ till the second repeated instance of $x_{0}$,\\
        $\vdots$
        \item $W_{n+1}$ is the subwalk beginning from the $n$th repeated instance of $x_{0}$ till the end of $W$.
    \end{itemize}
    Clearly, each of the above subwalks is a circuit. Since $W$ satisfies $\overline\Xi(W) < 0$, there is at least one $i$, $i\in\{1,2,\cdots,n+1\}$ such that $\overline\Xi(W_{i}) < 0$. Consider the $W_{i}$ from above that satisfies $\overline\Xi(W_{i}) < 0$. In case there are more than one such $W_{i}$'s, we select the one with the least number of vertices that appear more than once. If the selected $W_{i}$ is a cycle, we stop; else we proceed to Step 2.\\
    \textsf{Step 2}. Pick a vertex $x_{i}$, $i\neq 0$ that appears more than once in $W_{i}$. Consider the subwalks
    \begin{itemize}[label = $\circ$, leftmargin = *]
        \item $W_{i}^{(1)}$ obtained by concatenating $x_{0},(x_{0},x_{1}),x_{1},\cdots,$\\$x_{i-1},(x_{i-1},x_{i}),x_{i}$ (beginning from the initial vertex of $W_{i}$ and ending at the first instance of $x_{i}$), and $x_{i},(x_{i},x_{i+1'}),x_{i+1'},\cdots,x_{0}$ (beginning from the last instance of $x_{i}$ and ending at the final vertex of $W_{i}$);
        \item $W_{i}^{(2)}$ beginning from the first instance of $x_{i}$ and ending at the last instance of $x_{i}$.
    \end{itemize}
    Since $W_{i}$ satisfies $\overline\Xi(W_{i}) < 0$, one of the following is true: i. $\overline\Xi(W_{i}^{(1)}) < 0$, ii. $\overline\Xi(W_{i}^{(2)}) < 0$, iii. both $\overline\xi(W_{i}^{(1)})$ and $\overline\xi(W_{i}^{(2)}) < 0$.\\
    \emph{Recursive Step}: In case i. or ii. holds, we select the subwalk $W_{i}^{(j)}$, $j\in\{1,2\}$ that satisfies $\overline\Xi(W_{i}^{(j)}) < 0$. In case of the last one, we select the subwalk $W_{i}^{(j)}$ with less number of vertices, which appear more than once. If the selected $W_{i}^{(j)}$ is a cycle, we stop; else we generate $W_{i}^{(j)(1)}$ and $W_{i}^{(j)(2)}$ by the same procedure explained in Step 2 (Basis Step) and continue till we obtain a cycle.

    Armed with the above claim, consider the case when a closed contractive circuit $W$ satisfies $\overline\Xi(W) < 0$ is decomposed into $p$ cycles --- $W_{1},W_{2},\cdots,W_{p}$. Now, since $\overline\Xi(W) < 0$, at least one of the $W_{i}$'s, $i\in\{1,2,\cdots,p\}$ satisfies $\overline\Xi(W_{i}) < 0$. A cycle $W_{i}$ satisfying $\overline\Xi(W_{i})$ is the one that we are looking for. It contradicts our assumption that there is no cycle on $G$ that is contractive.

    Consequently, i) $\Rightarrow$ iii).

    iii) $\Rightarrow$ ii) A contractive cycle $W$ is a contractive circuit. If not, then there is at least one edge in $W$ that is repeated. But then the corresponding vertices are also repeated, which contradicts the definition of a cycle.

    ii) $\Rightarrow$ i) By definition, a contractive circuit is a closed contractive walk.

    The implication iii) $\Rightarrow$ i) follows at once.

    The last claim follows at once from Lemma \ref{lem:closedwalk}.\hfill{}$\qed$

    {\bf Proof of Lemma \ref{lem:cyclelem}}: Let $W' = j_{0},(j_{0},j_{1}),j_{1},\cdots,j_{k-1},$\\$(j_{k-1},j_{k}),j_{k},(j_{k},j_{i}),j_{i}$ be a walk obtained from Algorithm \ref{algo:cyclealgo}. Consider the sub-walk \(W = j_{i},(j_{i},j_{i+1}),j_{i+1}, \cdots\) \(j_{k-1}, (j_{k-1},j_{k}), j_{k}, (j_{k},j_{i}), j_{i},\)
             which is a cycle by construction. By of Algorithm \ref{algo:cyclealgo} all the vertices of $W$ are in $\P_{S}$. We claim that $\abs{W} \geq \lfloor\Phi(\abs{\P_{S}})\rfloor$. Assume, if possible, $\abs{W} < \lfloor\Phi(\abs{\P_{S}})\rfloor$. But
            \begin{align*}
               \abs{W} &= \abs{j_{i},(j_{i},j_{i+1}),j_{i+1},\cdots,j_{k-1},(j_{k-1},j_{k}),j_{k}} +\abs{j_{k},(j_{k},j_{i}),j_{i}} \\
                &= \abs{j_{i},(j_{i},j_{i+1}),j_{i+1},\cdots,j_{k-1},(j_{k-1},j_{k}),j_{k}} + 1.
            \end{align*}
            By hypothesis, $d_{\P_{S}}^{+} \geq \lfloor\Phi(\abs{\P_{S}})\rfloor$, which implies that
            \begin{align}
            \label{e:cyclepfi}
                \abs{N_{\P_{S}}^{+}} \geq \lfloor\Phi(\abs{\P_{S}})\rfloor.
            \end{align}
            By choice of $j_{i}$ in Algorithm \ref{algo:cyclealgo},
            \begin{align}
            \label{e:cyclepfii}
                \{j_{0},j_{1},\cdots,j_{i-1}\} \notin N_{\P_{S}}^{+}(j_{k}).
            \end{align}
            From \eqref{e:cyclepfi} and \eqref{e:cyclepfii}, it follows that $\abs{\{j_{i},j_{i+1},\cdots,j_{k}\}} \geq \abs{N_{\P_{S}}^{+}(j_{k})}$. But $\displaystyle{\abs{N_{\P_{S}}^{+}} \geq \lfloor\Phi(\abs{\P_{S}})\rfloor}$, which implies that\\ $\displaystyle{\abs{\{j_{i},j_{i+1},\cdots,j_{k}\}}\geq\lfloor\Phi(\abs{\P_{S}})\rfloor}$, and it is a contradiction. Consequently, $\abs{W} \geq \lfloor\Phi(\abs{\P_{S}})\rfloor + 1$.\hfill{}$\qed$

    {\bf Proof of Theorem \ref{t:mainres2}}:
        Since the given weighted digraph $G(\P,E(\P))$ is nicely connected, by Lemma \ref{lem:cyclelem} there exists cycle on $G(\P,E(\P))$ with all vertices of the cycle being in $\P_{S}$ and the length of the cycle is at least $\lfloor\Phi(\abs{\P_{S}})\rfloor$. Such a cycle can be detected by Algorithm \ref{algo:cyclealgo}.

            Consider a cycle $W = j_{0}, (j_{0},j_{1}), j_{1}, \cdots, j_{n-1}, (j_{n-1},$\\$j_{0}), j_{0}$ of length exactly $\lfloor\Phi(\abs{\P_{S}})\rfloor = n$ (say). Since $\{j_{0},j_{1},\cdots,j_{n-1}\}\in\P_{S}$, \eqref{e:negativeclosed} can be written as: $\sum_{k=1}^{n}w(j_{k-1},j_{k}) - \sum_{k=0}^{n}w(j_{k}) < 0$. Let
            \begin{align}
            \label{e:Xndefn}
                X_{n} \Let \sum_{k=1}^{n}w(j_{k-1},j_{k}) - \sum_{k=0}^{n}w(j_{k}).
            \end{align}
            Define the filtration \((\mathfrak F_m)_{m=0}^n\) by
            \begin{align*}
                \mathfrak{F}_{m} = \sigma\bigl\{w(j_{k-1},j_{k}),w(j_{\ell})\,\big|\, k = 0,\cdots,m-1,\:\ell = 0,\cdots,m\bigr\}.
            \end{align*}
            Since $G$ is nicely weighted,
            \begin{align}
                \EE^{\mathfrak{F}_{k-1}}[X_{k}] &= X_{k-1} + \EE^{\mathfrak{F}_{k-1}}[w(j_{k-1},j_{k}) - w(j_{k})]\nonumber\\
                &= X_{k-1} + \alpha - \beta\nonumber\\
                &< X_{k-1}\:\:\text{since $\alpha < \beta$ by Definition \ref{d:niceconnwt}.}\label{e:secpf1}
            \end{align}
            Let $(X_{m})_{m=0}^{n} \Let (\xi_{0} + M_{m} + A_{m})_{m=0}^{n}$ denote the a.s.\ unique Doob decomposition \cite[Theorem 5.2.10]{durrett} of the process $(X_{m})_{m=0}^{n}$. In other words, with \(M_0 \Let 0\) and \(A_0 \Let 0\), we have
			\[
			\left\{
			\begin{aligned}
				& \xi_{0}\Let X_{0},\\
				& M_{m}\Let\sum_{k=1}^{m}\bigl(X_{k}-\EE^{\mathfrak{F}_{k-1}}[X_{k}]\bigr),\\
				& A_{m} \Let \sum_{k=1}^{m}\bigl(\EE^{\mathfrak{F}_{k-1}}[X_{k}] - X_{k-1}\bigr),
			\end{aligned}
			\right.\qquad m = 1,\cdots,n,
			\]
			The inequality \eqref{e:secpf1} shows that $(X_{k})_{k=0}^{n}$ is an $(\mathfrak{F}_{k})_{k=0}^{n}$ strict supermartingale; the compensator process $(A_{k})_{k=0}^{n}$ is, therefore, strictly decreasing.

            The definition of $\xi_{0}$ shows that $\xi_{0} \leq 0$, and from \eqref{e:secpf1} we get $A_{n} \leq (\alpha - \beta)n$. Since
            \begin{align*}
                \PP(X_{n}>0) &= \PP(\xi_{0}+M_{n}+A_{n} > 0) \leq \PP(M_{n} + A_{n} > 0)\\
                &\leq \PP(M_{n} > -n(\alpha-\beta))\\
                &= \PP\biggl(\frac{M_{n}}{A+B} > \frac{-n(\alpha-\beta)}{A+B}\biggr),
            \end{align*}
            we apply Azuma's inequality \cite[Theorem 7.2.1]{AlonSp} to the zero-mean martingale process $(M_{m})_{m=0}^{n}$ to get
            \begin{align*}
                \PP(X_{n} > 0) &\leq \PP\biggl(\frac{M_{n}}{A+B} > \Bigl(\frac{-(\alpha-\beta)\sqrt{n}}{A+B}\Bigr)\sqrt{n}\biggr)\\
                &\leq \exp\biggl(-\frac{1}{2}\biggl(\frac{(\alpha-\beta)\sqrt{n}}{A+B}\biggr)^{2}\biggr),
            \end{align*}
            which gives the estimate in the theorem. The final assertion follows at once from Lemma \ref{lem:closedwalk}.\hfill{}$\qed$


\begin{thebibliography}{10}

\bibitem{Liberzon12}
A.~A. Agrachev, Y.~Baryshnikov, and D.~Liberzon.
\newblock On robust {L}ie-algebraic stability conditions for switched linear
  systems.
\newblock {\em Systems Control Lett.}, 61(2):347--353, 2012.

\bibitem{AlonSp}
N.~Alon and J.~H. Spencer.
\newblock {\em The probabilistic method}.
\newblock Wiley-Interscience Series in Discrete Mathematics and Optimization.
  John Wiley \& Sons, Inc., Hoboken, NJ, third edition, 2008.
\newblock With an appendix on the life and work of Paul Erd{\H{o}}s.

\bibitem{Bliman_IFAC}
P.~A. Bliman and G.~Ferrari-Trecate.
\newblock Stability analysis of discrete-time switched systems through lyapunov
  functions with nonminimal state.
\newblock {\em IFAC Conference on the Analysis and Design of Hybrid Systems
  (ADHS03), St. Malo, France}, 2003.

\bibitem{Bollobas}
B.~Bollob{\'a}s.
\newblock {\em Modern graph theory}, volume 184 of {\em Graduate Texts in
  Mathematics}.
\newblock Springer-Verlag, New York, 1998.

\bibitem{Cormen_algo}
T.~H. Cormen, C.~E. Leiserson, R.~L. Rivest, and C.~Stein.
\newblock {\em Introduction to algorithms}.
\newblock MIT Press, Cambridge, MA, third edition, 2009.

\bibitem{Daafouz02}
J.~Daafouz, P.~Riedinger, and C.~Iung.
\newblock Stability analysis and control synthesis for switched systems: a
  switched {L}yapunov function approach.
\newblock {\em IEEE Trans. Automat. Control}, 47(11):1883--1887, 2002.

\bibitem{durrett}
R.~Durrett.
\newblock {\em Probability: theory and examples}.
\newblock Cambridge Series in Statistical and Probabilistic Mathematics.
  Cambridge University Press, Cambridge, fourth edition, 2010.

\bibitem{Harris}
J.~M. Harris, J.~L. Hirst, and M.~J. Mossinghoff.
\newblock {\em Combinatorics and graph theory}.
\newblock Undergraduate Texts in Mathematics. Springer, New York, second
  edition, 2008.

\bibitem{Heemels_survey}
W.~P. M.~H. Heemels, B.~De~Schutter, J.~Lunze, and M.~Lazar.
\newblock Stability analysis and controller synthesis for hybrid dynamical
  systems.
\newblock {\em Philos. Trans. R. Soc. Lond. Ser. A Math. Phys. Eng. Sci.},
  368(1930):4937--4960, 2010.

\bibitem{HespanhaMorse}
J.~P. Hespanha and A.~S. Morse.
\newblock Stability of switched systems with average dwell-time.
\newblock In {\em Proc. of the 38th Conf. on Decision and Contr.}, pages
  2655--2660, Dec 1999.

\bibitem{Basar05}
H.~Ishii, T.~Ba{\c{s}}ar, and R.~Tempo.
\newblock Randomized algorithms for synthesis of switching rules for multimodal
  systems.
\newblock {\em IEEE Trans. Automat. Control}, 50(6):754--767, 2005.

\bibitem{knc_hscc14}
A.~Kundu and D.~Chatterjee.
\newblock Stabilizing discrete-time switched linear systems.
\newblock Proceedings of the 17th ACM International Conference on Hybrid
  Systems: Computation \& Control, 2014, Berlin, Germany, pp. 11-20.

\bibitem{abc}
A.~Kundu and D.~Chatterjee.
\newblock Stabilizing switching signals for switched systems.
\newblock To appear in IEEE Transactions on Automatic Control, doi:
  10.1109/TAC.2014.2335291.

\bibitem{Lee005}
J.~W. Lee and G.~E. Dullerud.
\newblock Uniform stabilization of discrete-time switched and markovian jump
  linear systems.
\newblock {\em Automatica}, 42:205--218, 2006.

\bibitem{Lee_07}
J.~W. Lee and G.~E. Dullerud.
\newblock Uniformly stabilizing sets of switching squences for switched linear
  systems.
\newblock {\em IEEE Transactions on Automatic Control}, 52:868--874, 2007.

\bibitem{Lee_Khar09}
J.~W. Lee and P.~P. Khargonekar.
\newblock Detectability and stabilizability of discrete-time switched linear
  systems.
\newblock {\em IEEE Transactions on Automatic Control}, 54:424--437, 2009.

\bibitem{Lewandowski10}
S.~Lewandowski.
\newblock Shortest paths and negative cycle detection in graph with negative
  weights {I}. the {B}ellman-{F}ord-{M}oore algorithm revisited.
\newblock {\em Technical Report 2010/05, Universit$\ddot{a}$t Stuttgart, FMI,
  Stuttgart, Germany}, 2010.

\bibitem{Liberzon}
D.~Liberzon.
\newblock {\em Switching in systems and control}.
\newblock Systems \& Control: Foundations \& Applications. Birkh\"auser Boston
  Inc., Boston, MA, 2003.

\bibitem{Antsaklis_survey}
H.~Lin and P.~J. Antsaklis.
\newblock Stability and stabilizability of switched linear systems: a survey of
  recent results.
\newblock {\em IEEE Trans. Automat. Control}, 54(2):308--322, 2009.

\bibitem{Lofberg04}
J.~L{\"o}fberg.
\newblock Y{A}{L}{M}{I}{P} : a toolbox for modeling and optimization in matlab.
\newblock {\em In Proceedings of IEEE International Symposium on Computer Aided
  Control Systems Design}.

\bibitem{mitra_ADT}
S.~Mitra, N.~Lynch, and D.~Liberzon.
\newblock Verifying average dwell time by solving optimization problems.
\newblock In {\em Hybrid systems: computation and control}, volume 3927 of {\em
  Lecture Notes in Comput. Sci.}, pages 476--490. Springer, Berlin, 2006.

\bibitem{papa_optimization}
C.~H. Papadimitriou and K.~Steiglitz.
\newblock {\em Combinatorial optimization: algorithms and complexity}.
\newblock Dover Publications Inc., Mineola, NY, 1998.
\newblock Corrected reprint of the 1982 original.

\bibitem{Shorten_review}
R.~Shorten, F.~Wirth, O.~Mason, K.~Wulff, and C.~King.
\newblock Stability criteria for switched and hybrid systems.
\newblock {\em SIAM Rev.}, 49(4):545--592, 2007.

\bibitem{SDPT3}
K.~C. Toh, M.~J. Todd, and R.~H. T{\"u}t{\"u}nc{\"u}.
\newblock On the implementation and usage of {SDPT}3---a {M}atlab software
  package for semidefinite-quadratic-linear programming, version 4.0.
\newblock In {\em Handbook on semidefinite, conic and polynomial optimization},
  volume 166 of {\em Internat. Ser. Oper. Res. Management Sci.}, pages
  715--754. Springer, New York, 2012.

\bibitem{allnegcycles}
T.~Yamada and H.~Kinoshita.
\newblock Finding all the negative cycles in a directed graph.
\newblock {\em Discrete Appl. Math.}, 118(3):279--291, 2002.

\bibitem{Zaroliagis08}
C.~Zaroliagis.
\newblock Negative cycles in weighted digraphs.
\newblock {\em Enclycopedia of Algorithms}, pages 576--578, 2008.

\bibitem{Zhai002}
G.~Zhai, H.~Bo, K.~Yasuda, and A.N. Michel.
\newblock Qualitative analysis of discrete-time switched systems.
\newblock {\em Proc. of the American Control Conference}, pages 1880--1885,
  2002.

\end{thebibliography}

\end{document}